\documentstyle[12pt]{article}

\begin{document}

\title{\bf The Spin Gap in the Context\\
of the Boson-Fermion Model\\
for High $T_c$ Superconductivity}

\author{{\bf J.~Ranninger and J.-M.~Robin}
\\{\small\it Centre de Recherches sur les 
Tr\`es Basses Temp\'eratures, }
\\{\small\it associ\'e \`a l'UJF, 
CNRS, BP 166, 38042 Grenoble-Cedex 9, France}}
\maketitle
\begin{abstract}
The issue of the spin gap in the magnetic susceptibility 
$\chi''({\bf q},\omega)$ in high $T_c$ superconductors is discussed 
within a scenario of a mixture of localized tightly bound electron 
pairs in singlet states (bi-polarons) and itinerant electrons. 
Due to a local exchange between the two species 
of charge carriers, antiferromagnetic correlations 
are induced amongst the itinerant electrons in the 
vicinity of the sites containing the bound electron pairs. 
As the temperature is lowered these exchange processes become 
spatially correlated leading to a spin wave-like spectrum in 
the subsystem of the itinerant electrons. 
The onset of such coherence is accompanied by the opening 
of a pseudo gap in the density of states of the electron subsystem 
whose temperature dependence is reflected in that of $\chi''
({\bf q},\omega)$ near ${\bf q}=(\pi,\pi)$ where a ``spin gap'' 
is observed by inelastic neutron scattering and NMR.
\end{abstract}
$\;\;\;${\small {\bf keywords}~: Pseudo gap, 
Magnetic susceptibility, Neutron scattering.

\vspace{1cm}

\noindent
$\;\;\;${\bf PACS Nb}~: 74.72.-h, 74.25.-q, 75.40Gb}

\newpage

The thermodynamic, magnetic and transport properties in underdoped samples 
of high $T_c$ superconductors (HT$_c$SC) show noticeable deviations from 
the usual Fermi liquid properties \cite{Battlogg} in the normal state 
phase above the superconducting transition temperature $T_c$ and 
below a characteristic temperature $T^\ast$. We have recently 
suggested \cite{Ranninger} that these effects might be due 
to the opening of a pseudo gap in the density of states (DOS) 
of the electrons.
Numerous experimental studies have led to the conclusion 
that possibly two types of charge carriers, 
almost localized ones and itinerant electrons, 
are involved in those materials \cite{carexp}, 
which can be phrased in terms of a simple model where free fermions 
on a lattice coexist with tightly bound localized electron pairs 
(such as bi-polarons) ---the Boson-Fermion model. 
As we have shown previously \cite{Ranninger2,Ranninger3}, 
such tightly bound localized electron pairs acquire itinerancy 
due to a precursor effect towards superconductivity, 
as the temperature is lowered and $T_c$ is approached. 
Such a scenario can potentially lead to very high values 
of $T_c$ since the superconductivity state is controlled by 
the Bose-Einstein condensation of the tightly bound electron 
pairs \cite{Ranninger4,Ranninger5}. 
The itinerancy of tightly bound electron pairs is achieved 
by resonating exchange processes between them and pairs of 
itinerant electrons and leads to the opening of a pseudo gap 
in the DOS in the subsystem of the itinerant electrons which, 
close to the Fermi energy, show strong deviations 
from Fermi liquid behaviour. 
The signature of this pseudo gap on the specific heat, 
the optical conductivity and the NMR relaxation rate have 
been recently studied by us in details \cite{Ranninger}.
A further eminent effect in the temperature interval 
$[T_c,T^\ast]$ involves the magnetic correlations in the 
electronic subsystem which is the topic of this short note.
One of the puzzling features of high $T_c$ compounds which has 
received considerable attention from experimental as well as 
theoretical studies is related to the so-called ``spin gap'' 
observed in the magnetic susceptibility $\chi''({\bf q},\omega)$ 
for the antiferromagnetic wave vector ${\bf q}=[\pi,\pi]$ 
in the metallic CuO$_2$ planes of these materials. 
Inelastic neutron scattering studies show a gap in the spin wave 
excitation spectrum at a frequency $E_G$ which opens up as soon 
as these materials are doped from the insulating into the metallic 
regime and increases in size upon further doping, until the optimally 
doped regime is reached \cite{Rossat}. As a function of temperature 
this ``spin gap'' is characterized by a transferral of spectral weight 
from frequencies above $E_G$ to below as the temperature increases 
from $T_c$ towards $T^\ast$ without that $E_G$ would noticeably change. 
Upon increasing the doping further through the so-called optimally 
doped into the overdoped situation, these spin gap features in $\chi''
({\bf q},\omega)$ rapidly disappear for very small amounts of additional 
doping. 
This indicates that a major part of the magnetic correlations 
have disappeared \cite{Bourges}  under the influence of this 
extra minute doping while $T_c$ is hardly at all affected as 
compared to its value of the optimally doped case. 
There is experimental evidence that the ``spin gap'' given by 
$E_G$ and the pseudo gap controlled by $T^\ast$ are essentially 
unrelated, given their contrasting dependence with doping which 
varies in opposite directions \cite{Rossat,Loram}. 
The present consensus is that the value of $E_G$ is essentially 
determined by band structure \cite{Bulut,Fukuyama} 
while the temperature dependence of the spin gap is controlled 
by $T^\ast$ which is largely determined by the onset of 
superconducting phase coherence \cite{Ranninger,Emery}. 
The closeness of the numerical values of $E_G$ and $T^\ast$ 
in the underdoped case may thus be completely fortuitous.
From the theoretical side this problem has received attention 
primarily from scenarios based on the strong electronic correlations 
of the CuO$_2$ planes in terms of the Hubbard model \cite{Bulut}, 
the $t-J$ model \cite{Fukuyama} and the some heuristic 
antiferromagnetic 2D-Heisenberg model \cite{Sokol}. 
The scenario based on the 2D correlated 
electron systems \cite{Bulut,Fukuyama} leads to a ``spin gap'' 
which is of kinematic origin, linked to nesting properties 
of the free electron dispersion. 
For a standard electron system having a Fermi surface centered 
around the $\Gamma$ point of the Brillouin zone, $E_G$ for the 
antiferromagnetic wave vector turns out to be given by $D-2\mu$ 
where $D$ denotes the bandwidth and $\mu$ the chemical potential 
measured from the bottom of the band. The evolution of the Fermi 
surface in HT$_c$SC as a function of doping is badly understood. 
The present experimental situation \cite{Marshall} suggests that 
for underdoped samples it consists of small closed ellipsoid like 
pockets centered roughly around 
$\left(\pm\frac{\pi}{2},\pm\frac{\pi}{2}\right)$ 
while approaching the optimally doped case it changes 
into a large Fermi surface, roughly given by that of a nearly 
half-filled quasi free tight binding model on a square lattice. 
Presently there is no unanimously accepted theoretical interpretation 
of this behaviour which is certainly linked to the strong correlations 
as well as charge transfer processes in the CuO$_2$ layers.

The purpose of the present study is to investigate the influence 
of the pseudo gap in the DOS of the fermionic subsystem upon 
the ``spin gap'' features seen by neutron scattering. 
We shall do this on the basis of the 2D Boson-Fermion 
mixture scenario for which the appearance of the pseudo gap and 
its manifestations in a number of physical quantities has been 
studied by us before \cite{Ranninger}.
The Boson-Fermion model is described by the Hamiltonian
\begin{eqnarray}
H & = & (zt-\mu)\sum_{i,\sigma}c^+_{i\sigma}c_{i\sigma}
-t\sum_{\langle i\neq j\rangle,\sigma}c^+_{i\sigma}c_{j\sigma}
+(\Delta_B-2\mu) \nonumber 
\\& & \times \sum_ib^+_ib_i
+v\sum_i [b^+_ic_{i\downarrow}c_{i\uparrow}
+c^+_{i\uparrow}c^+_{i\downarrow}b_i] 
\label{eq1}
\end{eqnarray}
where $c_{i\sigma}^{(+)}$ refers to the Fermion operators of the 
itinerant electrons in the metallic CuO$_2$ planes and $b^{(+)}_i$ 
refers to the Boson operators denoting the localized electron pairs 
in the dielectric layers separating the metallic planes. 
$i$ denotes some effective site involving 
adjacent molecular 
clusters of the metallic and 
dielectric planes \cite{Clusters}, 
spin indices are given by $\sigma$, the bare hopping integral 
for 
the electrons is given by $t$, the Boson energy level by $\Delta_B$ 
and the Boson-Fermion pair exchange coupling constant by $v$. 
In order to preserve charge conservations we impose a common 
chemical potential $\mu$. We have previously evaluated the Fermion and Boson 
one particle Green's function \cite{Ranninger2,Ranninger3} 
within a fully self-consistent lowest order diagramatic formulation 
amounting to solve the following set of non-linear equations for the 
Fermion and Bose single particle Green's function 
$G_F({\bf k},\omega_n)$, $G_B({\bf q},\omega_m)$ 
together with their corresponding self-energies 
$\sum_F({\bf k},\omega_n)$, 
$\sum_B({\bf q},\omega_m)$:
\begin{eqnarray}\begin{array}{lll}G_F({\bf k},\omega_n) 
& = & [i\omega_n-\epsilon_{\bf k}-\sum_F({\bf k},\omega_n)]^{-1},\\
\sum_F({\bf k},\omega_n) 
& = & \frac{1}{\beta}\;
\frac{v^2}{N}\displaystyle{\sum_{{\bf q},\omega_m}}
G_F(-{\bf k}+{\bf q},\omega_m-\omega_n) 
\\
& & \times G_B({\bf q},\omega_m), \\
& & 
\\
G_B({\bf q},\omega_m) & = & 
[i\omega_m-E_0-\sum_B({\bf q},\omega_m)]^{-1}, \\
\sum_B({\bf q},\omega_m) 
& = & 
-\frac{1}{\beta}\;\frac{v^2}{N}
\displaystyle{\sum_{{\bf k},\omega_n}}G_F(-{\bf k}
+{\bf q},-\omega_n+\omega_m) \\ 
& & 
\times G_F({\bf k},\omega_n). 
\label{eq2}
\end{array}
\end{eqnarray}
where $E_0=\Delta_B-2\mu$, $\epsilon_k=t(4-\sum_\delta 
e^{ik.\delta})-\mu$. 
$\beta$ denotes $1/k_BT$ and $N$ the total number of sites, 
while ${\bf k}$ and ${\bf q}$ refer to the wave vectors and 
$\omega_n$ and $\omega_m$ to the Matsubara frequencies of the Fermions 
and Bosons respectively. Using the previously numerically determined 
one particle Fermion Green's function we now evaluate the 
magnetic susceptibility
\begin{eqnarray}
\chi^"({\bf q},\omega) & = & \mu^2_B\mbox{ Im}\; G^R({\bf q},\omega) 
\nonumber \\
G^R({\bf q},t) 
& = & 
-\mu^2_B\frac{\theta(t)}{i\hbar}
\langle[S^-(q,t),S^+(-q,0)]\rangle , 
\label{eq3}
\end{eqnarray}
where 
$G^R({\bf q},\omega)$ 
denotes the Fourier transform of the retarded two-particle 
Green's function and
\begin{equation}
S^{+-}({\bf q},t)=\frac{1}{N}
\sum_ie^{i{\bf qr_i}}S_i^{+-}(t) 
\label{eq4}
\end{equation}
with 
$S^+_i=c^+_{i\uparrow}c_{i\downarrow}$ and $(S^+_i)^+=S^-_i$. 
Evaluating $G^R({\bf q},\omega)$ to lowest order 
(neglecting vertex corrections) we obtain:
\begin{equation}
G^R({\bf q},\omega_m)=
-\frac{1}{N}\frac{1}{\beta} \sum_{\bf k} \sum_{i\omega_n}
G_F({\bf k},i\omega_n)
G_F({\bf k}+{\bf q},i\omega_n+i\omega_m)
\label{eq5}
\end{equation}
which in the absence of Boson-Fermion exchange coupling yields 
the parti\-cularly simple form
\begin{equation}
\chi"({\bf q},\omega)=\pi\mu^2_B\frac{1}{N}
\sum_{\bf k}[f(\epsilon_{{\bf k+q}})
-f(\epsilon_{\bf k})]\delta(\omega+\epsilon_{\bf k}
-\epsilon_{\bf k+q}) 
\label{eq6}
\end{equation}
which for the antiferromagnetic {\bf q} vector 
${\bf q_A}=[\pi,\pi]$ and $T\rightarrow 0$ reduces to
\begin{equation}
\chi"({\bf 
q_A},\omega)=
\frac{\pi}{2}\rho(-\frac{\omega}{2}
-\mu+\frac{D}{2})\theta(\omega+2\mu-D) 
\label{eq7}
\end{equation}
where $D=8t$ and 
$\rho(\omega)=\frac{1}{N}\sum_{\bf k}\delta(\omega-\epsilon_k)$ 
is the DOS of the electrons.
From Equ.~\ref{eq7} we notice that the ``spin gap'' 
occurs at $\omega=D-2\mu$ which is the same result as 
that obtained in ref.~\cite{Bulut} and based on a $U>0$ Hubbard model 
(notice that in ref.~\cite{Bulut} the chemical potential is measured 
from the center of the band rather from the bottom as in our case). 
In our previous study of the interacting problem for a 2D system we 
have studied the evolution of the pseudo gap in the DOS of the single 
particle spectrum which occurs for $\omega\simeq 0$ showing an 
increase of spectral weight for $\omega$ slightly above and below 
$\omega\simeq 0$ in a region of width $2v$.  
Hence we expect from Equ. (\ref{eq7}) a peak in 
$\chi"({\bf q_A},\omega)$ slightly above the spin gap 
which grows as $T$ is diminished, while for higher frequencies 
$\chi"({\bf q_A},\omega)$ is essentially be given by the 
free DOS $\rho(-\frac{\omega}{2}-\mu+\frac{D}{2})$. 
In Fig.~1 we present the full numerical study of 
$\chi"({\bf q_A},\omega)$, given by Equ.~\ref{eq5}. 
We notice that as the temperature is lowered the ``spin gap'' 
becomes better and better defined with a slope of 
$\chi"({\bf q}_A,\omega)$ at $\omega=E_G\simeq D-2\mu$ which steadily 
increases as one approaches $T^\ast$ and then rapidly saturates 
as the temperature is further decreased below $T^\ast$. 
This behaviour of $\chi"({\bf q}_A,\omega)$ coincides with a 
Korringa type behaviour of $1/T_1T$ above $T^\ast$ and a noticeable 
deviation from it below $T^\ast$, which previously we have attributed 
to the opening of the pseudo gap in the DOS of the 
Fermions \cite{Ranninger}. The appearance of this pseudo gap 
has been found to be due to a destruction of well defined 
quasi particles in the vicinity of the Fermi level and the occurrence 
of a BCS-like spectrum involving several 
excitation branches \cite{Ranninger3}. 
As a consequence, similar to a BCS state, coherence effects play a 
dominant role as the temperature is decreased below $T^\ast$ and 
influence the magnetic susceptibility via the spectral one particle 
Fermion Green's functions entering the expression for 
$\chi"({\bf q},\omega)$ in Equ.~\ref{eq5}.

Our study of $\chi"({\bf q},\omega)$ in the entire 
Brillouin zone (Fig.~2) showed that apart from the region around 
the wave vector ${\bf q}_A$ there is another domain in $q$-space, 
corresponding to excitations of frequency $\sim\mu$, where the magnetic 
response is strongly influenced by coherence effects. 
This corresponds to wave vector ${\bf q}_B\simeq[\pm\frac{\pi}{4},\pm\frac{\pi}{4}]$ where the onset of 
coherence effects leads to the appearance of two well defined peaks 
in $\chi"({\bf q}_B,\omega)$ as compared to the case where Boson-two 
Fermion exchange processes are absent. 
This is particularly visible in the spectrum of the 1D Boson-Fermion model 
where our calculations are of better resolution (see Fig.~3). 
An experimental verification of this prediction of a strong 
temperature dependence of $\chi"({\bf q}_B,\omega)$ 
(apart from that already established for $\chi"({\bf q}_A,\omega)$) 
would shine new light on the origin of the peculiar spin dynamics 
of HT$_c$SC and in particular the two component scenario for HT$_c$SC.

The Boson-Fermion model discussed here describes a superconducting 
state controlled by phase rather than amplitude fluctuations with a 
mean field gap energy of the fermionic subsystem being much bigger  
than the energy of the phase fluctuations (see the discussion in ref.~7). 
Thus the Boson-Fermion model might be a reasonably good realization of 
the underdoped HT$_c$SC which can be considered as doped insulators 
and which, because of their low doping, have a very low superfluid 
density of the fermionic subsystem. In such a case phase fluctuations 
are expected to be determinant for the value of $T_c$ and 
the system's properties near $T_c$ are then essentially controlled 
by phase fluctuations \cite{Emery}.

Throughout this work we have assumed the following set of parameters 
$v=0.1$, $\Delta_B=0.4$, 
and $n_{tot}=\sum_{i,\sigma}
\langle c^+_{i\sigma}c_{i\sigma}\rangle
+2\sum_i\langle b^+_ib_i\rangle=1$. 
This choice corresponds to a small pocket Fermi surface centered around 
the $\Gamma$ point and with a $k_F\sim\pi/3$ \cite{Ranninger3}. 
Certainly in order to obtain a more realistic description of the 
``spin gap'' phenomenon, these parameters  not only will have to be 
modified, but also the Boson-Fermion model itself (Equ.~\ref{eq1}) 
would have to be generalized by explicitely distinguishing the fermionic 
sites from the bosonic ones and including correlations amongst the 
electrons which can yield the set of pockets of Fermi surfaces in 
fact observed experimentally. 
The present work, in complete analogy with previous work on this matter 
and involving electronic and magnetic correlations \cite{Bulut,Fukuyama}, 
cannot therefore be expected to provide reliable values for $E_G$ as 
measured by neutron scattering. It is however expected to correctly 
describe the temperature variation of $\chi''({\bf q},\omega)$ 
controlled by the same characteristic temperature $T^\ast$ at which 
the pseudo gap in the DOS of the electrons begins to open up and 
a superconducting phase coherence sets in 
above $T_c$.

\section*{Acknowledgements}
The authors would like to thank T.~Domanski 
and P.L.~Regnault for illuminating discussions.

\newpage

\begin{center}
{\Large \bf Figure Captions}
\end{center}

\noindent{\bf Figure 1}~: $\chi"({\bf q_A},\omega)$ as a 
function of frequency and for different temperatures 
(both in units of the bandwidth $D$), the steepest slope 
at $\omega$=$D$-$2\mu\sim$0.6 corresponding to $T=0.0085$. 
In the inset we present for comparison the temperature 
variation of the pseudo gap in the DOS of the Fermions, 
the doped pseudo gap corresponding to $T=0.0085$.

\vspace{1cm}

\noindent
{\bf Figure 2}~: 
$\chi"({\bf q},\omega)$ as a function of ${\bf q}$ over 
the entire Brillouin zone $q_x=q_y=[0,\pi]$ and $T=0.0085D$.

\vspace{1cm}

\noindent
{\bf Figure 3}~: $\chi"({\bf q},\omega)$ as a function of frequency 
for $T=0.001D$ for the 1D Boson-Fermion model 
and $q_x=q_y=[0,\pi/2]$.
\end{document}